% ------------------------------------------------------------------- 
% su.tex   Green's function for .....
% -------------------------------------------------------------------
% paper Keviczky-Saad-Hall
% 
% su.tex   [21 Jan 2004] [April 2003 su4.tex] [19 Jan 2004 su5.tex]
% [ 13 July 2004]
% -------------------------------------------------------------------
\def\ptitle{\tiny Green's function for a Schr\"odinger operator ...}
% -------------------------------------------------------------------
%  generic unix 12 fonts (lower case names) with no magstep
% --------------------------------------------------------------------
\font\tr=cmr12                          % Our default
\font\bf=cmbx12                         % Redefinition
                         % Redefinition
\font\it=cmti12                         % Redefinition
\font\trbig=cmbx12 scaled 1500          % Main Title
                          % Theorems                       
\font\tiny=cmr10                        % Running title
% --------------------------------------------------------------------
\output={\shipout\vbox{\makeheadline
                                      \ifnum\the\pageno>1 {\hrule}  \fi 
                                      {\pagebody}   
                                      \makefootline}
                   \advancepageno}

\headline{\noindent {\ifnum\the\pageno>1 
                                   {\tiny \ptitle\hfil
page~\the\pageno}\fi}}
\footline{}
% ---------------------------------------------------------------------

\tr 
%--------------------------------------------------------------------
\def\bra{{\rm <}}    % bra ket:  math mode (to replace angle)
\def\ket{{\rm >}}    %   ket  >
\def\nl{\hfil\break\noindent}  % new line after displayed equations
\def\ni{\noindent}             % noindent

\def\htab#1#2{{\hskip #1 in #2}}
\def\bra{{\rm <}} % bra < math mode
\def\ket{{\rm >}} % ket > math mode
\def\hi#1#2{$#1$\kern -2pt-#2} % hyphen \hi{N}{body} = N-body
\def\hy#1#2{#1-\kern -2pt$#2$} % hyphen hy{large}{N} = large-N
%--------------------------------------------------------------------
\def\dbox#1{\hbox{\vrule % Open box size 2#1 (Abrahams p 273) 
\vbox{\hrule \vskip #1\hbox{\hskip #1\vbox{\hsize=#1}\hskip #1}\vskip #1 
\hrule}\vrule}} 
\def\qed{\hfill \dbox{0.05true in}} % QED 
 % SQUARE 

%--------------------------------------------------------------------
% SPACING
% -------------------------------------------------------------------
\baselineskip 15 true pt  % draft 15 
\parskip=0pt plus 5pt 
\parindent 0.25in
\hsize 6.0 true in 
\hoffset 0.25 true in 
% 6 in width with 1.25 in margins default = (6.5, 0)
\emergencystretch=0.6 in                 % TEXBook p 107 : allows h-space 
\vfuzz 0.4 in                            % page-length flexibility
\hfuzz  0.4 in                           % line-length flexibility
\vglue 0.1true in
\mathsurround=2pt                        % Default is 2pt
\topskip=24pt                            % Default is 10pt
% ---------------------------------------------------------------------
%  References
% ---------------------------------------------------------------------
\newcount\zz  \zz=0  % switch for printing references
\newcount\q   %  reference number
\newcount\qq    \qq=0  % starting reference number-1   (usually zero)

\def\pref#1#2#3#4#5{\frenchspacing \global \advance \q by 1     % paper reference
    \edef#1{\the\q}{\ifnum \zz=1{\item{[{\the\q}]}{#2}{#3}~{ #4.}{~#5}\medskip} \fi}}

\def\prefm#1#2#3#4#5#6#7{\frenchspacing \global \advance \q by 1     % paper reference
    \edef#1{\the\q}{\ifnum \zz=1{\item{[{\the\q}]}{#2,~}{\it #3,~}{#4~}{\bf#5},
{#6.}{~#7}\medskip} \fi}}

\def\bref #1#2#3#4#5{\frenchspacing \global \advance \q by 1     % book reference
    \edef#1{\the\q}
    {\ifnum \zz=1 { %
       \item{[{\the\q}]} 
       {#2}, {#3} {(#4).}{~#5}\medskip} \fi}}

\def\gref #1#2{\frenchspacing \global \advance \q by 1  % general reference
    \edef#1{\the\q}
    {\ifnum \zz=1 { %
       \item{$^{\the\q}$} 
       {#2.}\medskip} \fi}}

\def\sref #1{[#1]}

 \def\srefs#1#2{{~[#1-#2]}}

\def\references#1{\zz=#1
   \parskip=2pt plus 1pt   % default is 0pt plus 1pt       
   {\ifnum \zz=1 {\noindent \bf References \medskip} \fi} \q=\qq
%--------------------------------------------------------------------
\pref{\hala}{R. Hall, N. Saad and A. von Keviczky, Spiked harmonic oscillators. J. Math. Phys. }{43}{(2002) 94-112} {}

\pref{\halb}{R. Hall, N. Saad and A. von Keviczky, Matrix elements for a generalized spiked harmonic oscilltor. J. Math. Phys. }{39}{(1998) 6345-6351}{}

\pref{\halc}{R. Hall and N. Saad, Variational analysis for a generalized spiked harmonic oscillator. J. Phys. A: Math. Gen. }{33}{(2000) 569-578}{}

\pref{\hald}{R. Hall, N. Saad and A. von Keviczky, Generalized spikes harmonic oscillator. J. Phys. A: Math. Gen. }{34}{(2001) 1169-1179}{}

\pref{\hale}{N. Saad, R. Hall and A. von Keviczky, Energy bounds for a class of singular potentials and some related series. J. Phys. A: Math. Gen. }{36}{(2003) 487-498}{}

\pref{\wats}{G. N. Watson, Notes on generating functions. I. Laguerre polynomials. J. London math. Soc. }{8}{(1933) 189-192}{}

\bref{\buch}{H. Buchholz}{The Confluent Hypergeometric Function}{New York: Springer-Verlag 1969}{}

\bref{\sriv}{H. M. Srivastava and H. L. Manocha}{A Treatise on Generating Functions}{New York: Halsted/Wiley 1984}{}
\bref{\elna}{Elna B. McBride}{Obtaining Generating Functions}{Springer, 1971}{}

\bref{\kam}{E. Kamke}{Differentialgleichungen - L\"osungsmethoden und L\"osungen}{B. G. Teubner, Stuttgart, 1983}{pp. 440, Formula Ia.}

\bref{\ince}{E. L. Ince}{Ordinary Differential Equations}{Dover Publications, N. Y., 1956}{}

\bref{\cod}{E. A. Coddington and N. Levinson}{Theory of Ordinary Differntial Equations}{McGraw-Hill, London, 1956}{}

\bref{\kamb}{E. Kamke}{Differentialgleichungen}{Gest \& Portig K.-G., Leipzig, 1962}{Vol. I.}

\bref{\wei}{ J. Weidmann}{Lineare Operatoren in Hilbertr\"aumen}{B. G. Teubner, Stuttgart, 1976 }{}

\bref{\rin}{ Riesz and Sz.-Nagy}{Functional Analysis}{Frederick Ungar, N. Y., 1965}{}

\bref{\hs} {H. Heuser}{Funktional Analysis}{B. G. Teubner, Stuttgart, 1986}{}

\bref{\nag}{ B. Sz.-Nagy}{Introduction to Real Functions and Orthogonal Expansions}{Oxford University Press, 1965}{}

\bref{\ws} {W. Schmeidler}{Integralgleichungen mit Anwendungen in Physik und Technik}{Akademische Verlagsgesellschaft Gest \& Portig K.-G., Leipzig, 1955}{p. 305, Theorem 58.}

\bref{\smi}{ F. Smithies}{Integral Equations}{Cambridge University Press, London, 1970}{}
}% end of ref list

\references{0}    % Initialization of reference numbers
% ------------------------------------------------------------------ end our ref.tex

% ----------------------------
% Preprint list
% ----------------------------
\htab{3.5}{CUQM-110}

\htab{3.5}{math-ph/0505070}

\htab{3.5}{May 2005}
%-------------------------------------------------------------------
%Title Page 
%-------------------------------------------------------------------
%-------------------------------------------------------------------
%Title Page 
%-------------------------------------------------------------------
\vskip 0.5 true in
%\centerline{\bf\trbig A Summation Formula via Green's Function }
%\centerline{\bf\trbig  in $L_2(0,\infty)$}
\centerline{\bf\trbig Green's function for a Schr\"odinger operator }
\vskip 0.1 true in
\centerline{\bf\trbig  and some related summation formulas}

\medskip
\vskip 0.25 true in
\centerline{Attila B. von Keviczky$^\dagger$, Nasser Saad$^\ddagger$ and Richard L. Hall$^\dagger$}
\bigskip
{\leftskip=0pt plus 1fil
\rightskip=0pt plus 1fil\parfillskip=0pt
\obeylines
$^\dagger$ Department of Mathematics and Statistics, Concordia University,
1455 de Maisonneuve Boulevard West, Montr\'eal, 
Qu\'ebec, Canada H3G 1M8.\par}

\medskip
{\leftskip=0pt plus 1fil
\rightskip=0pt plus 1fil\parfillskip=0pt
\obeylines
$^\ddagger$ Department of Mathematics and Statistics,
University of Prince Edward Island, 
550 University Avenue, Charlottetown, 
PEI, Canada C1A 4P3.\par}

\vskip 0.5 true in
%---------------------------------------------------------------------------
% Abstract
%---------------------------------------------------------------------------
\centerline{\bf Abstract}\medskip
\noindent Summation formulas are obtained for products of associated Lagurre polynomials by means of the Green's function $K$ for the Hamiltonian $H_0=-{d^2\over dx^2} + x^2 + Ax^{-2}
\quad (A>0).$  $K$ is constructed by an application of a Mercer type theorem that arises in connection with integral equations.  The new approach introduced in this paper may be useful for the construction of wider classes of generating function.
\bigskip
\nl{\bf Keywords}~~Schro\"odinger operators, singular potentials, Mercer's Theorem, Laguerre polynomials, Green's functions.
\bigskip
\nl{\bf PACS } 03.65.Ge
\vfil\eject

%---------------------------------------------------------------------------
% Preliminary Notions
%---------------------------------------------------------------------------
\ni{\bf 1. Introduction and main results}
\medskip
\noindent Since the early development of quantum mechanics, highly singular potentials have attracted much attention. Two main reasons for this are (1) regular perturbation theory can fail badly for such potentials, and (2) in physics one often encounters phenomenological potentials that are strongly singular at the origin such as certain types of nucleon-nucleon interaction, and singular models of fields in arbitrary dimensions.
A specific family of singular quantum Hamiltonians known as generalized harmonic oscillators given by
$$H(\lambda)=H_0+{{\lambda}\over{x^{\alpha}}}=-{d^2\over dx^2}+x^2+{A\over x^2}+{\lambda\over x^\alpha},\quad (A\geq 0,\ \alpha>0,\ \lambda \geq 0)\eqno(1)$$
and acting in the Hilbert space $L_2(0,\infty)$ have been subject to intensive investigation recently.  For a background and brief history of these problems we refer the reader to the summary in reference \sref{\hala}. We have shown \srefs{\hala}{\halb} that the set of eigenfunctions of $H_0 = H(0)$, namely 
$$
\psi_n(x) \equiv (-1)^n\sqrt{{2~(\gamma)_n\over n!\Gamma(\gamma)}}x^{\gamma-1/2}e^{-{x^2\over 2}} {}_1F_1(-n;\gamma;x^2)\hbox{ with }\gamma \equiv 1 + {1\over 2}\sqrt{1+4A}\quad (n = 0,1,2,\dots),\eqno(2)$$
constitutes an orthonormal basis for the Hilbert space $L_2(0,\infty)$. Here ${}_1F_1$ stands for the confluent hypergeometric function defined in terms of the associated Laguerre polynomials $L_n^{\gamma-1}(z)$ by  
$${}_1F_1(-n;\gamma;z)={n!\over (\gamma)_n}L_n^{\gamma-1}(z).\eqno(3)$$
This basis has proven to be useful in providing a complete variational study\srefs{\hala}{\hale} of the spectrum of $H(\lambda)$ for arbitrary fixed $A \geq 0,$ and $\lambda,\ \alpha > 0.$  The advantage over earlier studies in the Hermite basis $A = 0$ was that for $A > 0$ the \hi{H_0}{basis} itself derives from a singular problem with the term $A/x^2.$  In the present article, we explore another aspect of this basis. We shall prove that the eigenfunctions $\psi_n(x)$ satisfy the following identity:
$$
\sum_{n=0}^\infty {\psi_n(x)\psi_n(y)\over 4n + 2\gamma} = 
\cases{w(x)v(y)&for $0\leq y \leq x$\cr
\cr
v(x)w(y)&for $0\leq x \leq y$.\cr}\eqno(4)
$$
where
$$
w(x)v(y) = 2^{-1}~\sqrt{xy}~K_{\nu}\bigg({x^2\over 2}\bigg)I_{\nu}\bigg({y^2\over 2}\bigg),\quad\quad \hbox{where}\nu={1\over 2}(\gamma-1).
$$ 
In particular, we have, for $x=y$, that
$$
\sum_{n=0}^\infty  {\mid \psi_n(x) \mid^2 \over 4n+2\gamma}=2^{-1}~x~K_{\nu}\bigg({x^2\over 2}\bigg)I_{\nu}\bigg({x^2\over 2}\bigg).\eqno(5)
$$
Here $I_{\nu}(x)$ and $K_{\nu}(x)$ are modified Bessel functions of the first and second kind respectively.
There are direct applications for these idenities. An obvious application is that it can been seen as complimantary identity for Watson's famous result \sref{\wats} (see also \sref{\buch, p. 140, formula 14})
$$\sum\limits_{n=0}^\infty {(\gamma)_n\over n!(1+n)} {}_1F_1(-n;\gamma;x){}_1F_1(-n;\gamma;y)={}_1F_1(1;\gamma;y)\bigg[\Gamma(\gamma-1)~x^{1-\gamma}e^x-{1\over \gamma-1}{}_1F_1(1;\gamma;x)\bigg],\eqno(6)$$
valid for $x\geq y>0$. Second, it can be use in theory of coherent states to provide, for example, normalization factors of new class of coherent states labeled by confluent hypergeometric functions. Third, there are standard techniques known for
generating closed form sums for products of hypergeometric functions and related 
polynomials, such as Laguerre polynomials.  Srivastava {\it et al} \srefs{\sriv}{\elna} have discussed 
many different techniques that can be used for such 
purposes. It is noteworthy that the use of a kernel of a  
differential equation and a Mercer type theorem is not an idea that has been well 
explored in this context. It is our goal in the present article to 
show the usefulness of this approach to the construction of generating functions. 

\par In order to prove our main results, we organize the paper as follows. In Section 2, we introduce two linearly independent solutions of the second-order homogeneous differential equation $H_0u=[-{d^2\over dx^2} + (x^2 + Ax^{-2})] u = 0$. In Section 3, we construct the Green's function of $H_0$ and study some of its properties. The majorization of the Kernel operator $K(x,y)$ is investigated in Section 4. In Section 5, we introduce and prove a Mercer type theorem that allows us to conclude the absolute and uniformal convergence of the kernel $K(x,y)$ on the Hilbert space $L_2(0,\infty),$ and consequently prove our main results Theorem~1 and Theorem~2 from which formulas (4) and (5) follows immediately.

%---------------------------------------------------------------------------
% Second-order differential equation and its solutions
%---------------------------------------------------------------------------
\medskip
\ni{\bf 2. Second-order differential equation and its solutions}
\bigskip
\noindent If we set $u(z)=z^\alpha \psi(z)$ with $z={1\over 2}x^2$, we can easily show $H_0u=0$ reduces to
$${d^2\psi\over dz^2}+{(2\alpha+{1\over 2})\over z}{d\psi\over dz}-\bigg[1+{4\alpha(\alpha-{1\over 2})-A\over 4 z^2}\bigg]\psi(z)=0.\eqno(1)$$
If we adjust $\alpha$ so that $\psi(z)$ satifies the modified Bessel function
$${d^2\psi\over dz^2}+z^{-1}{d\psi\over dz}-\bigg[\nu^2z^{-2}+1\bigg]\psi(z)=0,$$
we obtain, for $\alpha={1\over 4}$ and $\nu ={1\over 4}\sqrt{1+4A}$, and from the basis $\{\sqrt{x}I_\nu({x^2\over 2}),\sqrt{x}K_\nu({x^2\over 2})\},$ the two linearly independent solutions 
$$v(x)=B\sqrt{x}I_\nu({x^2\over 2}),\quad w(x)=C\sqrt{x}K_\nu({x^2\over 2})\eqno(2),$$
where $B$ and $C$ are constants to be determined. We note that, the Wronskian of $I_\nu(z)$ and $K_\nu(z)$ satisfies \sref{\wats, p. 80, formula (90)}
$$I_\nu(z){K_\nu}^\prime(z)-{I_\nu}^\prime(z)K_\nu(z)=-z^{-1}, $$
which equation we divide by ${I_\nu}^2(z)$ and thereby obtain the derivative of the quotient $(K_\nu / I_\nu)(z)$, namely 
$$\bigg({K_\nu\over I_\nu}\bigg)^\prime(z)=z^{-1}{I_\nu}^{-2}(z).$$
We integrate this expression from $z$ to $\infty$ (using the properties $K_\nu(\infty)=0$ and $I_\nu(\infty)=\infty)$ and thus we arrive at
$$ \bigg({K_\nu\over I_\nu}\bigg)(z)= \int_z^\infty {1 \over \xi {I_\nu}^2(\xi)}d\xi.\eqno(3)$$ 
On the other hand, we make the replacements $z=x^2/2$, $K_\nu(x^2/2)=C^{-1}x^{-1/2}w(x)$ and $I_\nu(x^2/2)=B^{-1}x^{-1/2}v(x)$ in the immediately-preceeding formula, and find
$${K_\nu(x^2/2)\over I_\nu(x^2/2)}={B\over C}\times{w(x)\over v(x)}=\int_{x^2/2}^\infty {1\over \xi{I_\nu}^2(\xi)}d\xi,$$
where in the last integral we make the substitution $I_\nu(\xi)=B^{-1}({2\xi})^{-1/4}v(\sqrt{2\xi})$, thereby giving us 
$${B\over C}\times{w(x)\over v(x)}=\int_{\xi=x^2/2}^\infty \xi^{-1}\bigg[{v(\sqrt{2\xi})\over B(\sqrt{2\xi})} \bigg]^{-2}d\xi=2B^2\int_{\xi=x^2/2}^\infty {1\over 2\xi v^2(\sqrt{2\xi})(2\xi)^{-1/2}}d\xi= 2B^2\int_{r=x}^\infty {1\over v^2(r)}dr.$$
\nl The last integral expression was obtained via the substitution $r=\sqrt{2\xi}$. Equating the first expression with the last in the above equations leads to 
$${w(x)\over v(x)}= 2BC\int_{r=x}^\infty {1\over v^2(r)}dr\hbox{ or equivalently }w(x)=2BCv(x)\int_{r=x}^\infty {1\over v^2(r)}dr.\eqno(4)$$
However, reduction of the order of the original differential equation  implies that 
$$v(x)\int_{r=x}^\infty {1\over v^2(r)}dr$$ 
is the other solution of $H_0u=0,$ independent of $v(x)$, which result lets us conclude that $2BC=1$ or $BC=1/2$.

\medskip
%---------------------------------------------------------------------------
% Mapping Properties of the Green's Function in $L_2(0,\infty)$
%---------------------------------------------------------------------------
\ni{\bf 3. Mapping Properties of the Green's Function in $L_2(0,\infty)$}
\medskip
\noindent For the operator
$$H_0=-{d^2\over dx^2} + (x^2 + Ax^{-2}),\eqno(1)$$ 
the linear space $D(H_0)$, consisting of all functions $u \in C^2(0,\infty)\cap C[0,\infty)$ with $u(0)=0$, becomes a normed linear space by setting $\parallel u \parallel_\infty \equiv sup\mid u([0,\infty))\mid$. For the Green's function of the linear transformation $H_0:D(H_0)\mapsto C[0,\infty)$, we now formulate a well known result from the theory of ordinary differential equations, which may be easily arrived at from the properties of Green's function as stated in\srefs{\ince}{\kamb}. 

\bigskip
% ----------------------------------------------
\noindent{\bf Lemma 1.}
% -----------------------------------------------
 {\it The differential operator $H_0$ maps $D(H_0)$ bijectively onto $C[0,\infty)$  by means of ${H_0}^{-1}:f\mapsto u$, where
$$
({H_0}^{-1})f(x)=u(x)\equiv w(x)\int_{\xi=0}^x v(\xi)f(\xi)d\xi + v(x)\int_{\xi=x}^\infty w(\xi)f(\xi)d\xi,\eqno(2)
$$
with $\parallel {H_0}^{-1}f \parallel_\infty \leq 4A^{-1/2}\parallel f \parallel_\infty$.}
\medskip
\noindent{\bf Proof:} The boundedness of ${H_0}^{-1}$ is simply a consequence of the inequality $$v(x)\int_{\xi=x}^\infty [v(\xi)]^{-2}d\xi \leq {\sqrt{2}\over a}[v(x)]^{-1}\eqno(3)$$
where $a\equiv \root{4} \of{A},$ the value of $x$ where the potential $q_A(x)=x^2+Ax^{-2}$ assumes its minimum. Indeed, since the function $q_A(x)$ assumes its minimum value $q_A(\root{4} \of{A})=2\sqrt{A} = 2a^2$ at the point $x=a\equiv \root{4} \of{A}$, we note that 
$$
v(x) = v(x') + (x-x')v'(x') + \int_{\xi=x'}^x (x-\xi)~q_A(\xi)~v(\xi)~d\xi\quad \hbox{ for }\quad 0\leq x'
\leq x,
$$ 
and from this directly deduce   
$$
v(x) \geq v(x') + 2a^2\int_{\xi=x'}^x (x-\xi)~v(\xi)~d\xi\quad\hbox{ for }\quad 0\leq x'\leq x.
$$
By substituting $v(\xi) \geq v(x') + 2a^2\int_{\eta=x'}^\xi (\xi-\eta)v(\eta)d\eta$ into the previous integral inequality, we further arrive at 
$$
v(x) \geq \bigg\{1 +{1\over 2!}2a^2(x-x')^2 \bigg\}v(x') + {1\over 3!}(2a^2)^2\int_{\xi=x'}^x (x-\xi)^3~v(\xi)~d\xi\quad\hbox{ for }\quad 0\leq x'\leq x.
$$
Iterating this procedure leads to $v(x) \geq v(x')\cosh \sqrt{2}a(x-x')$ for $x \geq x'\geq 0.$ In particular 
$$
v(x) \geq 2^{-1}\exp(\sqrt{2}a(x-x'))v(x')\hbox{ for }x\geq x'\hbox{ and }
v(x) \geq 2^{-1}\exp(\sqrt{2}a(x-a))v(a)\hbox{ for }x\geq a,
$$ 
which can also be written in the form $v(x') \leq 2\exp(-\sqrt{2}a(x-x'))v(x)$ for $x \geq x'\geq 0$ and this in turn leads to 
$$
\int_{\xi =a}^x v(\xi)d\xi \leq {\sqrt{2}\over a}v(x).
$$
This result is also evident from the integration of the inequality $v(\xi) \leq 2\exp(-\sqrt{2}a(x-\xi))v(x)$ with respect to $\xi$ on the interval $(a,x)$. As a result of $v(\xi) \geq 2^{-1}\exp(\sqrt{2}a(\xi - x))v(x)$ for all $\xi \geq x$, or by means of $[v(\xi)]^{-2} \leq 4\exp(2\sqrt{2}a(x - \xi))[v(x)]^{-2}$, leads to the inequality
$$
\int_{\xi=x}^\infty [v(\xi)]^{-2}d\xi \leq 4\int_{\xi=x}^\infty \exp(2\sqrt{2}a(x - \xi))d\xi \times[v(x)]^{-2}\leq {\sqrt{2}\over a}[v(x)]^{-2},$$
which proves (3.3). The injective nature of ${H_0}^{-1}:C[0,\infty)\mapsto D(H_0)$ is demonstrated as follows. Let $u(x)$ be a solution of $H_0u =0$ with $u(0)=0$, whence $u(x)=Bw(x) + Cv(x)$. On account of the asymptotic behavior of $w(x)$ and $v(x)$ as $x\to 0^+$, namely $w(x)$ and $v(x)\to \infty$ and $0$ respectively as $x\to 0^+$, combined with $\parallel u \parallel_\infty < \infty$ leads to $B=0,$ that is to say $u(x)= Cv(x)$; however, the asymptotic behavior of $v(x)$, namely $$v(x) = {1\over \sqrt{\pi x}}e^{x^2\over 2}[1+O(x^2)]$$ as $x \to \infty$, implies that $C=0$. Thus $H_0:D(H_0)\mapsto C[0,\infty)$ is bijective. This completes the proof.\qed
\medskip
We have therefore that the Green's function $K(x,y)$ of the differential operator $H_0:D(H_0)\mapsto C[0,\infty)$, which permits us to write the action of ${H_0}^{-1}$ in terms of integration on $[0,\infty)$, has the integral representation
$$
{H_0}^{-1}f(x)= \int\limits_0^\infty K(x,y)~f(y)~dy,\eqno(4)$$
\nl where
$$
K(x,y)=\cases{w(x)v(y),&for $0\leq y\leq x$,\cr
\cr
	v(x)w(y),&for $0\leq x\leq y$.\cr}\eqno(5)
$$
\nl It thus becomes evident that the kernel $K(x,y)$ is a continuous non-negative function on $[0,\infty)\times [0,\infty)$ because, for $\nu={1\over 2}(\gamma-1)={1\over 4}\sqrt{1+4A},$ $\sqrt{x}I_{\nu}(x^2/2)$ and $\sqrt{x}K_{\nu}(x^2/2)$ are continuous on $[0,\infty)$ and $(0,\infty)$ respectively, and, furthermore, both are positive on $(0,\infty)$. Thus $K(x,y)$ is continuous on $[0,\infty)\times [0,\infty) \setminus \{(0,0)\}$. 

For the continuity at $(0,0)$ we again turn to the asymptotic behavior of $w(x)$ and $v(x)$ as $x\to 0^+$. In the lower sector $\{(x,y):0\leq \hbox{arctg}(y/x)\leq \pi/4\}$ of the first quadrant of the $(x,y)$-plane, we have that 
$$\eqalign{
K(x,y)&= w(x)v(y)\cr
&= {1\over 4\pi\nu}x^{1/2 -2\nu}
\big [1 + O(x^{\epsilon})\big ]y^{1/2 + 2\nu}\big [1+O(y^4)\big ]\cr
&={1\over 4\pi\nu}(xy)^{1/2}(y/x)^{2\nu}\big [1+O(y^4)\big ]\big [1 + O(x^{\epsilon})\big ]\cr
&\to 0\hbox{ as }(x,y)\to (0,0),\cr}
$$  
because in this sector $\mid y/x\mid\  \leq 1$. Consequently, $K(x,y)$ is continuous in the lower  sector $\{(x,y):0\leq \hbox{arctg}(y/x)\leq \pi/4\}$, whereas the symmetry of $K(x,y)$ - i. e. $K(x,y)= K(y,x)$ - guarantees the continuity in upper sector $\{(x,y):\pi/4 \leq \hbox{arctg}(y/x)\leq \pi/2 \}$ of the first quadrant of the $(x,y)$-plane. 

We know further that $H_0\equiv -d^2u/dx^2 + [x^2 + Ax^{-2}]u = 0\quad(A > 0)$ is a symmetric lower semi-bounded operator in the Hilbert space $L_2(0,\infty)$ with domain of definition consisting of all $L_2(0,\infty)$-functions $u$ vanishing at $0$, having absolutely continuous derivative $u'\in L_2(0,\infty)$ on $[0,\infty)$ such that $[x^2 + Ax^{-2}]u(x)$ is also an $L_2(0,\infty)$-function in $x$. Since the linear manifold $C_0^\infty(0,\infty)$, of all complex valued infinitely differentiable function on $(0,\infty)$ with compact support, lies dense in this domain of definition of $H_0$ as well as in $L_2(0,\infty)$, we readily conclude that the domain of definition of $H_0$ is also a dense subset of $L_2(0,\infty)$. Thus $H_0$ possesses a Friedrichs' extension, which extension we again denote by $H_0$, and this extension \sref{\wei, Sec. 7.2-7.3}-{\sref{\rin, p. 335} is a self-adjoint operator  in $L_2(0,\infty)$.  For $H_0\psi_n=E_n\psi_n$ and because the orthonormalized set of eigenfunctions (1.2)
with corresponding eigenvalues $E_n = 4n+2\gamma = 4n + 2 + \sqrt{1+4A}$ of $H_0$ forms \sref{\hala} a complete orthonormal set of functions of the Hilbert space $L_2(0,\infty)$, we shall have that the spectrum of this Friedrichs' extension $H_0$ is a purely point-spectrum, consisting only of the simple eigenvalues $E_n= 4n+2\gamma = 4n + 2 + \sqrt{1+4A}$. Thus the spectral family $\{P_\mu: \mu \in \Re\}$ of $H_0$, which is an ``increasing" projection-operator valued Saltus function on the set $\Re$ of real numbers \sref{\nag, p. 92}-\sref{\rin, Ch. I, Sec. 7}, is
$$P_\mu\equiv 
\cases{0,&for  $-\infty<\mu<2\gamma$\cr
\cr
\psi_0\otimes \psi_0,&for $4(0)+2\gamma\leq \mu <4(1)+2\gamma$\cr
\cr
\psi_0\otimes \psi_0 + \psi_1\otimes \psi_1,&for $4(1)+2\gamma\leq \mu <4(2)+2\gamma$
\cr\cr
\psi_0\otimes \psi_0 + \psi_1\otimes \psi_1+\psi_2\otimes \psi_2,&for $4(2)+2\gamma\leq \mu <4(3)+2\gamma$\cr
\cr
\dots\cr
\cr
\sum\limits_{k=0}^{n-1} \psi_k\otimes \psi_k,& for $4(n-1)+2\gamma\leq \mu <4(n)+2\gamma$\cr
\cr
\dots
\cr}\eqno(6)
$$
For any two $L_2(0,\infty)$-functions $\psi$ and $\phi$, the expression $\psi\otimes \phi$ denotes the operator of rank $1$ defined by  
$$
(\phi\otimes \psi)f(x)\equiv \bra f \mid \psi \ket \phi(x) = \bigg[\int_{\xi=0}^\infty f(\xi) \overline{\psi(\xi)}d\xi \bigg]\times\phi(x) =\int_{\xi=o}^\infty (\phi\otimes \psi)(x,\xi)f(\xi)d\xi,\eqno(7)
$$
with $L_2\hbox{-kernel }(\phi\otimes \psi)(x,y) \equiv \phi(x)\overline{\psi(y)}$ on $L_2(0,\infty)$, which in our case turns out be an integral operator. It is further clear that the spectral family $\{P_\mu: \mu \in \Re\}$ is increasing in $\mu$ - i. e. $P_\lambda \leq P_\mu$ for $\lambda \leq \mu$ - as well as continuous from the right - i. e. $P_\mu = P_{\mu+0}$ - in the sense of strong convergence (denoted by $\to$) in the Hilbert space $L_2(0,\infty).$ Moreover, $P_\mu \to 0$ or $I$ (the identity operator) according as $\mu \to -\infty$ or $\infty$. In consequence, the spectral decomposition \sref{\rin , Sec. 120, p. 320} of our self-adjoint operator $H_0$ allows $H_0$ to be represented (as well as functions of $H_0$) as a Stieltjes integral of $\mu$ (functions of $\mu$) with respect to the spectral family $\{P_\mu\}$ on the set $\Re =(-\infty,\infty)$ of real numbers. This turns out to be
$$H_0 = \int_{-\infty}^\infty \mu d_\mu P_\mu \eqno(8)$$
with
$$D(H_0) = \left\{f\in L_2(0,\infty):\int_{-\infty}^{\infty}\mu^2d_{\mu}\|P_{\mu}f\|^{2} =
 \sum_{n=0}^{\infty}(4n+2\gamma)^2|\bra f|\psi_{n}\ket|^2 < \infty\right\}$$
and
$${H_0}^{-1} = \int_{-\infty}^\infty \mu^{-1} d_\mu P_\mu =\sum_{n=0}^\infty (4n+2\gamma)^{-1}[P_{4n+2\gamma}-P_{4n+2\gamma-0}] =\sum_{n=0}^\infty (4n+2\gamma)^{-1}(\psi_n\otimes \psi_n),   
\eqno(9)$$
where $P_{4n+2\gamma}-P_{4n+2\gamma-0}= \psi_n\otimes \psi_n$ is the projection onto the eigenspace $(L.H.)(\psi_n)$ spanned by the single eigenfunction $\psi_n$ for each of the integers $n \geq 0$. The abbreviation $(L.H.)$ stands for ``the linear hull of whatever is between the two brackets to its immediate right", and `linear hull' means the set of all linear combinations. Consequently, the Green's function $K(x,y)$ of the differential operator $H_0$, namely the kernel of our self-adjoint operator $H_0$ restricted to the previous domain $D(H_0)$, takes on the form
$$
K(x,y)= \sum_{n=0}^\infty (4n+2\gamma)^{-1}(\psi_n\otimes \psi_n)(x,y)=\sum_{n=0}^\infty {\psi_n(x)\psi_n(y)\over 4n+2\gamma} \hbox{ a. e. on }[0,\infty)\times [0,\infty)   
\eqno(10)$$
with respect to Lebesgue measure on $[0,\infty)^2 = [0,\infty)\times [0,\infty)$, which shall turn out to be a positive $L_2$-kernel on $[0,\infty)$ with ``finite double norm" \sref{\smi , p. 13}. Herein we must emphasize the almost everywhere (a. e.) nature of the immediately-preceeding equality. The finite double norm of kernel $K(x,y)$ is defined as
$$
\mid\mid\mid K \mid\mid\mid  \equiv \sqrt{\int_0^\infty\int_0^\infty \mid K(x,y)\mid^2dydx}=\sqrt{\sum_{n=0}^\infty (4n+2\gamma)^{-2}} < \infty, \eqno(10)
$$
and consequently many of the ideas, but not all, that led to of Mercer's Theorem \sref{\smi , p. 127} are applicable. However, because $\sum_{n=0}^\infty (4n+2\gamma)^{-1}= \infty$, it cannot be expected that all of the results of Mercer's Theorem carry over; specifically, $K$ as an operator ``on" the Hilbert space $L_2(0,\infty)$ fails to be an operator of trace class. 
%---------------------------------------------------------------------------
% Majorization Property of the Kernel of Operator $K$ on $L_2(0,\infty)$
%---------------------------------------------------------------------------
\medskip
\noindent {\bf 4. Majorization Property of the Kernel of Operator $K$ on $L_2(0,\infty)$}
\medskip
\noindent Let us now call the extension of $H_0$ to all of $L_2(0,\infty)$ the operator $K$, which is an operator ``on" $L_2(0,\infty)$ (instead of {\it in} $L_2(0,\infty)$). We may do this, on 
account \sref{\smi , Th. 4.5.1, p. 63} of the fact that kernel $K(x,y)$ is an $L_2$-kernel  and therefore the norm  
relation 
$$
\parallel K \parallel \leq \mid\mid\mid K \mid\mid\mid \equiv \sqrt{\int_0^\infty\int_0^\infty \mid K(x,y)\mid^2dydx}=\sqrt{\sum_{n=0}^\infty (4n+2\gamma)^{-2}} < \infty,\eqno(1)
$$
guarantees that operator $K$ has domain of definition $D(K)=L_2(0,\infty)$. Herein, $
\parallel K \parallel$ and $\mid\mid\mid K \mid\mid\mid$  denote the operator norm of $K$ and double norm of its kernel $K(x,y)$ respectively. The action of $K$ on $L_2(0,\infty)$ is
$$
(Kf)(x) = \int_0^\infty K(x,y)f(y)dy = \sum_{n=0}^\infty {\bra f |\psi_n\ket \over 4n+2\gamma}\psi_n{(x)}\hbox{ a. e. in }x\hbox{ on }[0,\infty), \eqno(2)
$$ 
and moreover, this operator $K$ on $L_2(0,\infty)$ is also the Friedrichs' extension of $H_0$. We first consider the kernel $K(x,y)$ as the Green's function of $H_0$, and note that the completeness of the orthonormal basis $\{\psi_n: n \geq 0\}$, where the normalized eigenfunction $\psi_n$ corresponds to the simple eigenvalue $\lambda_n =4n+2\gamma$, entails that
$$
\bra Kf \mid g\ket = \sum_{n=0}^\infty {\bra f \mid \psi_n \ket \bra \psi_n \mid  g\ket\over 4n+2\gamma}\quad\hbox{ for all }f\hbox{ and }g\in L_2(0,\infty).\eqno(3)
$$
In particular, we replace $f$ and $g$ herein by the sequence of $L_2(0,\infty)$-functions $\delta_n(x)$, tending weakly towards the Dirac-$\delta$ function $\delta(x)$, defined by 
$$
\delta_n(x)\equiv n\big[1-n\mid x \mid\big]\quad \hbox{ for }\mid x \mid\leq 1/n\hbox{ and }0\hbox{ otherwise on the set }\Re,
$$
where $\Re$ is the set of real numbers. It becomes immediately clear, that out of $\bra Kf \mid f \ket$ always exceeding each of the finite sums $\sum_{k=0}^N [4n+2\gamma]^{-1}\mid \bra f \mid \psi_k \ket \mid^2$, the inequality
$$
\bra K\delta_n(\cdot -x) \mid \delta_n(\cdot -x) \ket \geq \sum_{k=0}^N {\bra \delta_n(\cdot -x) \mid \psi_n\ket\bra \psi_n \mid \delta_n(\cdot -x)\ket\over 4n+2\gamma}\hbox{ for all }N \geq 0
$$
implies, by way of letting $n\to \infty$ and holding $x\geq 0$ fixed, that 
$$K(x,x) \geq \sum_{k=0}^N [4n+2\gamma]^{-1}\mid \psi_k(x)\mid^2\quad\hbox{ for all } N\geq 0.$$ 
Therefore, it follows that
$$
\sum_{n=0}^\infty {\mid \psi_n(x)\mid^2\over 4n+2\gamma} \leq K(x,x)\hbox{ for all }x\in[0,\infty),
$$
where the function $K(x,x)$ has the precise form  given by ($\nu={1\over 4}\sqrt{1+4A}$)

$$K(x,x) = w(x)v(x)=2^{-1}~x~I_{\nu}\bigg({x^2\over 2}\bigg)K_{\nu}\bigg({x^2\over 2}\bigg)\eqno(4)
$$
as well as asymptotic behavior as $x \to 0^+$ and $\infty$ respectively given by
$$K(x,x)=\cases{{1\over \sqrt{1+4A}}~x~\big[1+O(x^\epsilon)],&for $x\to 0^+$\cr
\cr
2^{-1}~x^{-1}~\big[1+O(x^{-2})],&for  $x\to \infty$.\cr}\eqno(5)
$$
\bigskip
%-----------------------------------------------------------------------------------------
% Mercer Theorem Type Properties of the Kernel $K(x,y)$ of the Operator $K$
%-----------------------------------------------------------------------------------------
\ni{\bf 5. Mercer Theorem Type Properties of the Kernel $K(x,y)$}
\medskip
\noindent From the asymptotic behavior (4.5) of the majorizing function $K(x,x)$ of $\sum_{n=0}^\infty [4n+2\gamma]^{-1}\mid \psi_n(x)\mid^2$ as well as the continuity of $K(x,x)$ on $[0,\infty)$, we readily see that if we let $M_\infty \equiv \sup \{K(x,x): x \in [0,\infty)\}$, the above asymptotic behavior guarantees that $M_\infty < \infty$, then 
$$
\sum_{n=0}^\infty {\mid \psi_n(x)\mid^2\over 4n+2\gamma} \leq M_\infty\hbox{ for all }x\in[0,\infty).\eqno(1)
$$
It therefore follows that $\sum_{n=0}^\infty [4n+2\gamma]^{-1}\mid \psi_n(x)\mid^2$ converges for all $x\in [0,\infty)$. Moreover, the Cauchy-Schwarz inequality allows us to write for all non-negative integers $N$ that 
$$
\sum_{n=N}^\infty{\mid \psi_n(x)\psi_n(y)\mid\over 4n+2\gamma}\leq \big \{\sum_{n=N}^\infty {\mid \psi_n(x)\mid^2\over 4n+2\gamma}\big \}^{1/2}\times  \big \{\sum_{n=N}^\infty {\mid \psi_n(y)\mid^2\over 4n+2\gamma}\big \}^{1/2}\hbox{ for all }x,y \in [0,\infty).\eqno(2)
$$ 
It is further follows that the series $\sum_{n=0}^\infty[4n+2\gamma]^{-1}\psi_n(x)\psi_n(y)$ converges absolutely, because out of the immediately-preceeding inequality shall follow 
$$\eqalign{
&\sum_{n=N}^\infty{\mid \psi_n(x)\psi_n(y)\mid\over 4n+2\gamma}\leq \big \{\sum_{n=N}^\infty {\mid \psi_n(x)\mid^2\over 4n+2\gamma}\big \}^{1/2}\times \sqrt{M_\infty}\hbox{ for all }x\in [0,\infty)\hbox{ and}\cr
& \sum_{n=N}^\infty{\mid \psi_n(x)\psi_n(y)\mid\over 4n+2\gamma}\leq \big \{\sum_{n=N}^\infty {\mid \psi_n(y)\mid^2\over 4n+2\gamma}\big \}^{1/2}\times \sqrt{M_\infty}\hbox{ for all }y\in [0,\infty).\cr}\eqno(3)
$$
\nl Let $\epsilon > 0$. It then follows that, for every $y\in [0,\infty)$, there exists an integer $N=N(y)$ such that 
$$
\sum_{n=N}^\infty{\mid \psi_n(x)\psi_n(y)\mid\over 4n+2\gamma}\leq \epsilon \sqrt{M_\infty}\eqno(4)$$
\nl for all $x\in [0,\infty)$, provided $N = N(y)$ is chosen sufficiently large; 
and there exists correspondingly for every $x\in [0,\infty)$ an integer $N=N(y)$ such that
$$
\sum_{n=N}^\infty{\mid \psi_n(x)\psi_n(y)\mid\over 4n+2\gamma}\leq \epsilon \sqrt{M_\infty}\eqno(5)$$
for all $x\in [0,\infty)$, provided $N = N(x)$ is chosen sufficiently large. These two statements (5.4) and (5.5) are valid because the series $\sum_{n=0}^\infty [4n+2\gamma]^{-1}\mid \psi_n(x)\mid^2$ is majorized by the constant $M_\infty$ on $[0,\infty)$. Hence we are led to the following Mercer type theorem.
\medskip
\noindent{\bf Theorem 1.} {\it The kernel $K(x,y)$ of the operator $K$ on the Hilbert space $L_2(0,\infty)$ with spectral decomposition $K = \sum_{n=0}^\infty [4n + 2\gamma]^{-1}(\psi_n \otimes \psi_n)$  possesses the following property: the convergence of the series
$$
\sum_{n=0}^\infty {\psi_n(x)\psi_n(y)\over 4n+2\gamma} = K(x,y)\eqno(6)$$
is absolute and uniform on every compact subset of $[0,\infty)\times [0,\infty)$. }
\medskip
\noindent{\bf Proof:} The series $\sum_{n=0}^\infty[4n+2\gamma]^{-1}\mid \psi_n(x)\psi_n(y)\mid$ is uniformly convergent in $x$ on $[0,\infty)$ for every $y \geq 0$, as well as uniformly convergent in $y$ on $[0,\infty)$ for every $x \geq 0$, as evident from (5.4) and (5.5). Therefore $\sum_{n=0}^\infty[4n+2\gamma]^{-1} \psi_n(x)\psi_n(y)$  represents a continuous function in variable $x$ on $[0,\infty)$ for every $y \geq 0$, as well as a continuous function in variable $y$ on $[0,\infty)$ for every $x \geq 0$. For every $L_2(0,\infty)$-function $f$ the Fourier expansion of $Kf$ has the form
$$
(Kf)(x) = \int\limits_0^\infty K(x,y)f(y)dy
 = \sum_{n=0}^\infty {\bra f | \psi_n\ket \over 4n+2\gamma}\psi_n(x)\quad \hbox{ a. e. in }x\hbox{ on }[0,\infty)
$$
and possesses the following property
$$\eqalign{
\sum_{n=N}^\infty \mid {\bra f\ |\ \psi_n\ket\over 4n+2\gamma}\psi_n(x)\mid &\leq \bigg \{\sum_{n=N}^\infty \mid \bra f|
\psi_n\ket \mid^2\bigg \}^{1/2}\bigg \{\sum_{n=N}^\infty {\mid \psi_n(x)\mid^2 \over (4n+2\gamma)^2}\bigg \}^{1/2}\cr
&\leq \bigg \{\sum_{n=N}^\infty \mid \bra f | \psi_n\ket \mid^2\bigg \}^{1/2}\big \{\sum_{n=N}^\infty {\mid \psi_n(x)\mid^2 \over (4n+2\gamma)}\big \}^{1/2} \cr
&\leq \bigg \{\sum_{n=N}^\infty \mid \bra f | \psi_n\ket \mid^2\bigg \}^{1/2}\times \sqrt{M_\infty}, \cr}
$$
and therefore the sum $\sum_{n=0}^\infty [4n+2\gamma]^{-1}\bra f|\psi_n\ket \psi_n(x)$ is an absolutely and uniformly convergent series of continuous functions on $[0,\infty)$ whose limit is continuous on $[0,\infty)$, and hence 
$$
(Kf)(x) = \int_0^\infty K(x,y)f(y)dy = \sum_{n=0}^\infty {\bra f |\psi_n\ket\over 4n+2\gamma}\psi_n(x)\hbox{ for all }x\in[0,\infty).
$$
\nl Since in the series $\sum_{n=0}^\infty [4n+2\gamma]^{-1}(\psi_n \otimes \psi_n)$ the set of kernels $\{(\psi_n\otimes \psi_n):\ n \geq 0\}$ constitutes an orthonormal set of $L_2(0,\infty)$-kernels, the Riesz-Fischer Theorem, combined with $K(x,y)= \sum_{n=0}^\infty [4n+2\gamma]^{-1}(\psi_n\otimes \psi_n)(x,y)$ a. e. on $[0,\infty)\times [0,\infty)$ and the fact that $\sum_{n=0}^\infty [4n+2\gamma]^{-2} < \infty$, leads to 
$$
\mid\mid\mid K - \sum_{n=0}^\infty 
{\psi_n\otimes \psi_n \over 4n+2\gamma}\mid\mid\mid = \big \{\int_0^\infty \int_0^\infty |K(x,y)-\sum_{n=0}^\infty {\psi_n(x)\psi_n(y)\over 4n+2\gamma}|^2dydx\big \}^{1/2} = 0
$$ 
and furthermore to  
$$
\int_0^\infty \bigg[K(x,y) - \sum_{n=0}^\infty {\psi_n(x)\psi_n(y)\over 4n+2\gamma}\bigg]f(y)dy =0\hbox{ for all }f \in L_2(0, \infty).
$$ 
Now let ${\bf C}$ denote any compact interval $[a,b]$ contained in $[0,\infty)$, and consider the immediately-preceeding equality for all functions continuous on ${\bf C}$ and vanishing on $[0,\infty)\setminus {\bf C}$. Note that these functions belong to $L_2(0,\infty)$.  This in turn implies that 
$$
\int_{\bf C} \bigg[K(x,y) - \sum_{n=0}^\infty{\psi_n(x)\psi_n(y)\over 4n+2\gamma}\bigg]f(y)dy = 0$$
\nl for all $x\in {\bf C}$ and $f$ continuous on ${\bf C}$ with $f([0,\infty)\setminus {\bf C})=0$.  By choosing $f$ for any arbitrary, however momentarily, fixed $x\in {\bf C}$ as follows,
$$
f(y) = K(x,y) - \sum_{n=0}^\infty{\psi_n(x)\psi_n(y)\over 4n+2\gamma}\hbox{ and }f([0,\infty)\setminus {\bf C})\equiv 0, 
$$
where for this fixed $x\in {\bf C}$ the series represents a continuous function in variable $y$ on set ${\bf C}$ vanishing on $[0,\infty)\setminus {\bf C}$, and substituting it into the immediately above equality, we obtain that 
$$
\int_C \mid K(x,y) - \sum_{n=0}^\infty{\psi_n(x)\psi_n(y)\over 4n+2\gamma}\mid^2dy = 0\hbox{ for all }x\in {\bf C}.
$$
Because the integrand above is a continuous function of y on the compact subset ${\bf C}$, which is an arbitrary finite closed interval contained in $[0,\infty)$, we obtain
$$
K(x,y) - \sum_{n=0}^\infty{\psi_n(x)\psi_n(y)\over 4n+2\gamma} = 0\hbox{ for all }x,y\in [0,\infty), 
$$
\nl which, for $x=y\in [0,\infty),$ specifically yields 
$$
K(x,x) = \sum_{n=0}^\infty{\mid \psi_n(x)\mid^2\over 4n+2\gamma}\hbox{ for all }x\in [0,\infty).
$$
We now invoke Dini's Theorem, which states: Every monotone sequence of real valued continuous functions on a compact metric space with continuous limit, converges uniformly to its limit. Hence by Dini's Theorem \sref{\nag, p. 66} the convergence of $K(x,x) = \sum_{n=0}^\infty [4n+2\gamma]^{-1}\mid \psi_n(x)\mid^2$ is therefore uniform on every compact subset of $[0,\infty)$, and therefore the series $\sum_{n=0}^\infty [4n+2\gamma]^{-1}\psi_n(x)\psi_n(y)$ converges absolutely and uniformly on every compact subset of $[0,\infty)\times[0,\infty)$ with limit $K(x,y)$ - i. e. $K(x,y) = \sum_{n=0}^\infty [4n+2\gamma]^{-1}\psi_n(x)\psi_n(y)$ for all $(x,y)\in [0,\infty)\times [0,\infty)$. This completes the proof.\qed

\medskip

Again we note that this is only a Mercer type theorem, because it makes no conclusion about the operator $K$ or its $L_2(0,\infty)$-kernel $K(x,y)$ being of trace class; whereas Mercer's theorem makes an affirmative statement \sref{\rin, pp. 245-246}, see also \sref{\hs}-\sref{\smi}, concerning the trace class nature of operator $K$. As consequence of this Mercer type theorem for our Green's function $K(x,y)$, we return to the property of the spectral decomposition of the operator $K$ discussed before, because any complex valued function $W(\mu)$ of the real variable $\mu$ on $(-\infty,\infty)$ determines \sref{\rin, Sec. 127 and 128} an operator $W(K)$ {\it in} $L_2(0,\infty)$ defined by
$$
W(K) = \int_{-\infty}^\infty W(\mu)d_\mu P_\mu = \sum_{n=0}^\infty W(4n+2\gamma)[P_{4n+2\gamma}-P_{4n+2\gamma-0}] =\sum_{n=0}^\infty W(4n+2\gamma)(\psi_n\otimes \psi_n),
$$ 
whose domain of definition $D(W(K))$ consists of all $L_2(0,\infty)$-functions satisfying
$$
\int_{-\infty}^\infty \mid W(\mu)\mid^2d_\mu \bra P_\mu f|f\ket = \int_{-\infty}^\infty \mid W(\mu)\mid^2d_\mu \parallel P_\mu f \parallel^2 = \sum_{n=0}^\infty \mid W(4n + 2\gamma)\mid^2\mid \bra f |\psi_n\ket \mid^2 < \infty.
$$
It is clearly evident that the operators $W(K)$ are always densely defined, regardless of the function $W$ considered on $\Re=(-\infty,\infty)$, because every domain of definition $D(W(K))$ always contains $(L.H.)(\psi_n(n \geq 0))$. Thus we have that the inverse $[\lambda I - K]^{-1}$ of the operator $\lambda I - K$, in the normed algebra of bounded linear operators on  $L_2(0,\infty)$, comes about from the complex valued function $W(\mu)= [\lambda - \mu]^{-1}$ of the real variable $\mu$ and takes the form
$$\eqalign{
[\lambda I - K]^{-1} &= \int_{-\infty}^\infty [\lambda -\mu]^{-1}d_\mu P_\mu \cr
& = \sum_{n=0}^\infty [\lambda -(4n+2\gamma)]^{-1}[P_{4n+2\gamma}-P_{4n+2\gamma-0}]\cr &=\sum_{n=0}^\infty [\lambda -4n -2\gamma]^{-1}(\psi_n\otimes \psi_n),}
$$
provided $\lambda \notin \sigma(K) = \{4n+2\gamma: n\hbox{ a non-negative integer}\}$, namely the  spectrum of operator $K$ in the algebra of bounded linear operators on  $L_2(0,\infty)$, with continuous kernel 
$$
[\lambda - \cdot]^{-1}(x,y) = \sum_{n=0}^\infty{\psi_n(x)\psi_n(y)\over\lambda -4n-2\gamma}\hbox{ for all }(x,y)\in [0,\infty)\times [0,\infty),
$$
where the convergence is absolute on $[0,\infty)\times [0,\infty)$ and uniform on every compact subset of $[0,\infty)\times [0,\infty)$. We may consequently summarize our results in the following two theorems.
\medskip
\noindent{\bf Theorem 2.} {\it The $C[0,\infty)$-function $K(x,x)$ arising out of the Green's function of the differential operator $H_0=-{d^2\over dx^2} + x^2 + Ax^{-2}
\quad (A>0)$ satisfies:
$$
\sum_{n=0}^\infty  {\mid \psi_n(x) \mid^2 \over 4n+2\gamma}= K(x,x)  = w(x)v(x)=2^{-1}~x~I_{{1\over 2}(\gamma-1)}\bigg({x^2\over 2}\bigg)K_{{1\over 2}(\gamma-1)}\bigg({x^2\over 2}\bigg)
$$
where $\gamma=1+{1\over 2}\sqrt{1+4A},$ with uniform convergence on every compact subset of $[0,\infty).$  $K(x,x)$ has the asymptotic behaviour 
$$K(x,x)=\cases{{1\over \sqrt{1+4A}}~x~\big[1+O(x^\epsilon)],&for $x\to 0^+$\cr
\cr
2^{-1}~x^{-1}~\big[1+O(x^{-2})],&for  $x\to \infty$.\cr}
$$
}
\noindent{\bf Theorem 3.} {\it The continuous kernel $K(x,y)$ on $[0,\infty)\times [0,\infty)$, arising from the Green's function of the differential operator $H_0=-{d^2\over dx^2} + x^2 + Ax^{-2}
\quad (A>0)$, satisfies:
$$
K(x,y) = \sum_{n=0}^\infty {\psi_n(x)\psi_n(y)\over 4n + 2\gamma} = 
\cases{w(x)v(y)&for $0\leq y \leq x$\cr
\cr
v(x)w(y)&for $0\leq x \leq y$.\cr}
$$ on $[0,\infty)\times [0,\infty) $
with convergence being absolute and uniform on every compact subset of $[0,\infty)\times [0,\infty)$, where
$$
w(x)v(y) = 2^{-1}~\sqrt{xy}~K_{{1\over 2}(\gamma-1)}\bigg({x^2\over 2}\bigg)~I_{{1\over 2}(\gamma-1)}\bigg({y^2\over 2}\bigg).
$$}
We further conclude that, for the orthonormal basis $\{ \psi_n(x): n \geq 0\}$, (1.2), of the Hilbert space  $L_2(0,\infty)$, the two new summation formulas (1.4) and (1.5) follows immediately.
\medskip
\noindent {\bf Acknowledgment}
\medskip
\noindent Partial financial support of this work under Grant Nos. GP3438 and GP249507 from the 
Natural Sciences and Engineering Research Council of Canada is gratefully 
acknowledged by two of us (respectively [RLH] and [NS]).
\vfil\eject
%-------------------------------------
\references{1}
%-------------------------------------

\end